\documentclass[twocolumn,showpacs,preprintnumbers,amsmath,amssymb]{revtex4}
\usepackage{graphicx}% Include figure files
\usepackage{epsfig}
\usepackage{dcolumn}% Align table columns on decimal point
\usepackage{bm}% bold math

\begin{document}

\preprint{}
%%%%%%%%%%%%%%%%%%%%%%%%%%%%%%%%%%%%%%%%%%%%%%%%%%%
\title{Avoided Quantum Criticality in YBa$_2$Cu$_3$O$_y$ and La$_{2-x}$Sr$_x$CuO$_4$}

\author{J.E.~Sonier,$^{1,2,*}$ F.D.~Callaghan,$^{1}$ Y.~Ando,$^{3}$ R.F.~Kiefl,$^{2,4}$ 
J.H.~Brewer,$^{2,4}$ C.V.~Kaiser,$^{1}$ V.~Pacradouni,$^{1}$ S.-A. Sabok-Sayr,$^{1}$
X.F.~Sun,$^{3}$ S.~Komiya,$^{3}$ W.N.~Hardy,$^{2,4}$ D.A.~Bonn$^{2,4}$, and  R.~Liang$^{2,4}$}

\email{jsonier@sfu.ca}

\affiliation{$^1$Department of Physics, Simon Fraser University, Burnaby, British Columbia V5A 1S6, Canada \\
$^2$Canadian Institute for Advanced Research, 180 Dundas Street West, Toronto, Ontario M5G 1Z8, Canada \\
$^3$Central Research Institute of Electric Power Industry, Komae, Tokyo 201-8511, Japan\\
$^4$Department of Physics and Astronomy, University of British Columbia, Vancouver, British Columbia V6T 1Z1, Canada} 

\date{\today}
%%%%%%%%%%%%%%%%%%%%%%%%%%%%%%%%%%%%%%%%%%%%%%%%%%%%%%%

\begin{abstract}
Spin-glass (SG) magnetism confined to individual weakly interacting vortices 
is detected in two different families of high-transition-temperature ($T_c$)
superconductors, but only in samples on the low-doping side of the 
low-temperature normal state metal-to-insulator crossover (MIC). Our findings 
unravel the mystery of the MIC, but more importantly identify the true location 
of the field-induced quantum phase transition (QPT) in the superconducting 
(SC) state. The non-uniform appearance of magnetism in the vortex state favours 
a surprisingly exotic phase diagram, in which spatially inhomogeneous competing 
order is stabilized at the QPT, and an `avoided' quantum critical point (QCP) 
is realized at zero magnetic field.
\end{abstract}

\maketitle
For nearly two decades, arrival at a firm theory for high-$T_c$ superconductivity 
has been hindered by an incomplete characterization of 
the phase diagram for doped copper oxides. 
Zero-field (ZF) muon spin rotation ($\mu$SR) \cite{Panago:02,Panago:05} 
and neutron scattering \cite{Khaykovich:05} studies of magnetism in pure 
and Zn-doped La$_{2-x}$Sr$_x$CuO$_4$, point to the possible existence of a QCP under 
the SC `dome' --- corresponding to a zero-temperature phase transition at which a 
competing order is stabilized, and about which unusual properties emerge. 
The ZF-$\mu$SR measurements of Refs.~\cite{Panago:02,Panago:05}
support the occurrence of a QCP at what has been argued 
\cite{Tallon:01} to be a universal critical doping concentration $p_c \! \approx \! 0.19$, whereas
the neutron studies \cite{Khaykovich:05} suggest there is a QCP near
$p_c \! = \! 0.125$. 

%\twocolumn
\begin{figure*}
\centering
\includegraphics[width=19.0cm]{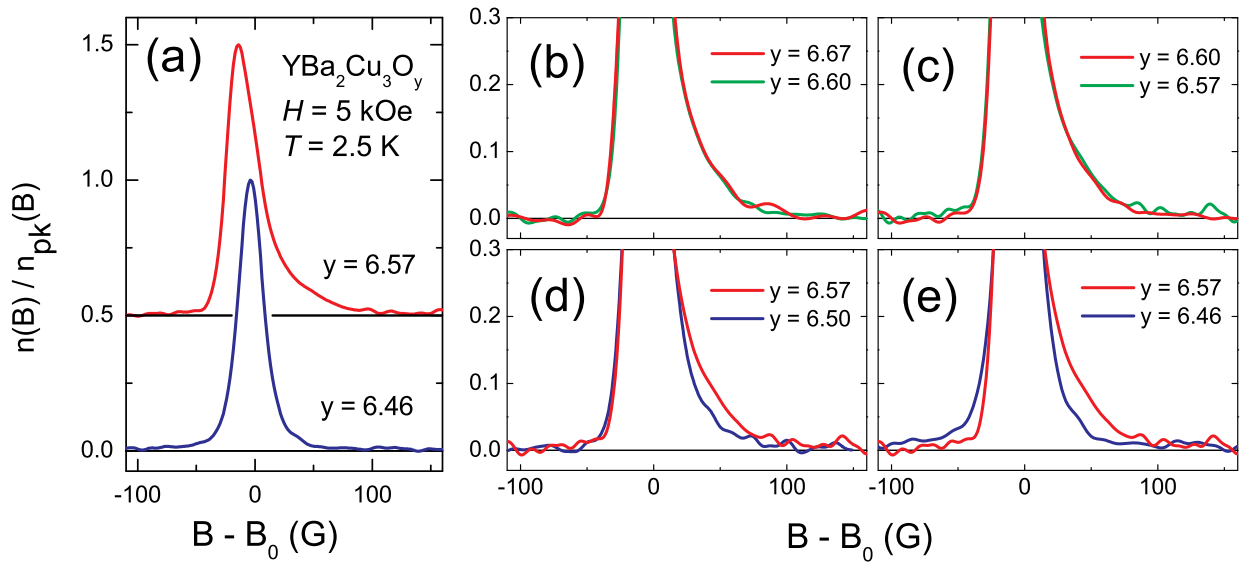}
\caption{(Color online) Doping dependence of the $\mu$SR line shapes for 
YBa$_2$Cu$_3$O$_y$ at $H \! = \! 5$~kOe and $T \! = \! 2.5$~K. 
(a) Examples of the full $\mu$SR line shape for samples above 
($y \! = \! 6.57$) and below ($y \! = \! 6.46$) the critical doping $y_c \! = \! 6.55$
for the normal-state MIC. (b), (c), (d), (e) Blowups of the `tail' region of the line shapes 
for samples above ($y \! = \! 6.67$, 6.60 and 6.57) and below ($y \! = \! 6.50$ 
and 6.46) $y_c \! = \! 6.55$. For comparison, all line shapes have been 
normalized to their respective peak amplitude $n_{\rm pk}(B)$. Furthermore, to account for 
changes in the magnetic penetration depth, the widths of the line shapes in (b), (c), (d), 
and (e) have been made equivalent by rescaling the horizontal $B \! - B_0$ axis, where
$B_0$ is the applied magnetic field.} 
\label{fig1}
\end{figure*}
%\onecolumn

It is now well established that a common intrinsic normal-state property of 
high-$T_c$ superconductors is the occurrence of a field-induced MIC at low 
temperatures, and at a non-universal doping 
\cite{Ando:95,Boebinger:96,Fournier:98,Ono:00,Sun:03,Hawthorn:03,Sun:04,Dagan:04,Dagan:05}.
Dagan {\it et al.} \cite{Dagan:04,Dagan:05} hypothesized that the MIC in 
electron-doped Pr$_{2-x}$Ce$_x$CuO$_4$ is associated 
with a QCP at $p_c \! \approx \! 0.165$, at which remnants of the antiferromagnetic 
(AF) phase disappear. However, measurements of the MIC do not imply there
is a competing magnetic order hidden in the SC phase at zero magnetic field. 
In fact, single-phase Pr$_{2-x}$Ce$_x$CuO$_4$ is difficult to grow, and 
magnetism is likely to reside in lightly doped regions of the sample
at $H \! = \! 0$. Likewise, one could never draw a firm conclusion from 
earlier experiments on hole-doped La$_{2-x}$Sr$_x$CuO$_4$, showing that 
an external magnetic field enhances static magnetism in underdoped samples 
\cite{Katano:00,Lake:02,Savici:05} and increases spin fluctuations in optimally 
or overdoped samples \cite{Lake:01}. More recently, spin-density-wave (SDW) order 
was detected by neutron scattering at $H \! > \! 30$~kOe in a 
La$_{1.856}$Sr$_{0.144}$CuO$_4$ sample that did not exhibit 
static magnetic order at zero field \cite{Khaykovich:05}. 
Combined with earlier works, the neutron results support a proposed phase 
diagram \cite{Demler:01} in which the pure superconductor undergoes a QPT to 
coexisting SC and SDW orders. However, neither the zero-field QCP deduced from the 
neutron studies, nor the QCP inferred from ZF-$\mu$SR experiments, correspond to 
the critical doping for the normal-state MIC. Kivelson {\it et al.} \cite{Kivelson:02} 
have proposed that the `true' field-induced QPT is one 
in which the competing order is stabilized in a halo around weakly interacting 
vortex lines. In this situation the magnitude of the competing order parameter is 
spatially inhomogeneous, and may only be detectable by a local probe technique.     

%\twocolumn
\begin{figure*}
\centering
\includegraphics[width=19.0cm]{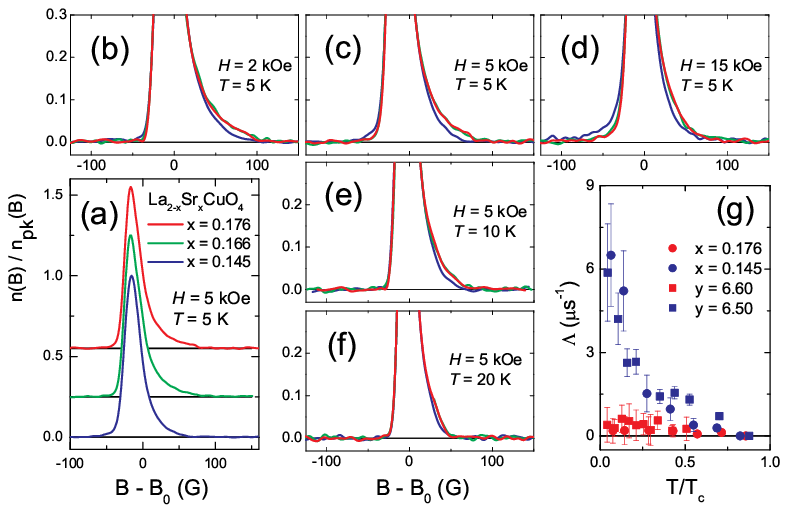}
\caption{(Color online) Doping, temperature and magnetic field dependences of the $\mu$SR line 
shapes for La$_{2-x}$Sr$_x$CuO$_4$. (a) Full $\mu$SR line shapes for samples 
above ($x \! = \! 0.176$ and $x \! = \! 0.166$) and below ($x \! = \! 0.145$) 
the critical doping $x_c \! \approx \! 0.16$ for the normal-state MIC. 
(b), (c), (d), (e), (f) Blowups of the 'tail' region of the $\mu$SR line shapes. 
The height and width of the line shapes have been normalized in the same way as 
in Fig.~1. Panels (b), (c) and (d) show the magnetic field dependence of the
line shapes, and panels (c), (e) and (f) show the temperature dependence.
(g) Temperature dependence of the relaxation rate 
$\Lambda$ for La$_{2-x}$Sr$_x$CuO$_4$ (circles) and YBa$_2$Cu$_3$O$_y$ (squares) 
at $H \! = \! 5$~kOe. The red and blue symbols denote samples on the high-doping 
and low-doping sides of the critical doping for the normal-state MIC, respectively.}
\label{fig2}
\end{figure*}
%\onecolumn    

For the present study we used weak magnetic fields applied perpendicular to the 
CuO$_2$ layers of La$_{2-x}$Sr$_x$CuO$_4$ and YBa$_2$Cu$_3$O$_y$ single crystals on 
either side of the critical dopings $x_c \! \approx \! 0.16$ 
\cite{Boebinger:96,Sun:03,Hawthorn:03} and $y_c \! \approx \! 6.55$ 
\cite{Sun:04} for the MIC, to locally suppress superconductivity by formation of 
a vortex lattice (VL). The vortex cores were 
probed using $\mu$SR spectroscopy (at TRIUMF,Canada), which is an extremely 
sensitive probe of local 
internal magnetic fields. Like a tiny bar magnet, the magnetic moment of a muon 
implanted in a sample precesses about the local magnetic field $B$ with an angular 
frequency $\omega_\mu \! = \! \gamma_\mu B$, where 
$\gamma_\mu \! = \! 0.0852$~$\mu$s$^{-1}$~G$^{-1}$ is the muon gyromagnetic ratio. 
By measuring the time evolution of the muon spin polarization $P(t)$ via the 
anisotropic distribution of decay positrons, the internal magnetic field 
distribution $n(B)$ of the sample is determined \cite{Sonier:00}. 
 ZF-$\mu$SR measurements at $T \! \geq \! 2.5$~K
indicate that none of our samples contain static electronic moments, which
is an essential requirement for establishing the presence of hidden
competing magnetic order. 

Figure~1 shows the `tail' regions of the Fourier transforms of $P(t)$ measured 
in the vortex state of YBa$_2$Cu$_3$O$_y$ near the critical doping 
$y_c \! \approx \! 6.55$. The Fourier transform of $P(t)$, 
often called the `$\mu$SR line shape', provides a fairly accurate visual 
illustration of $n(B)$ sensed by the muons. On the high-doping side of the MIC, 
the $\mu$SR line shapes for the $y \! = \! 6.67$, $y \! = \! 6.60$ and $y \! = \! 6.57$
samples are nearly identical, while those for the $y \! = \! 6.50$ and 
$y \! = \! 6.46$ samples below the MIC are clearly different. 
At $y \! = \! 6.50$, there is a clear suppression of the high-field tail, which 
corresponds to the spatial region of the vortex cores. A previous $\mu$SR study of 
$y \! = \! 6.50$ at higher fields concluded that the unusual high-field tail 
originates from AF vortex cores \cite{Miller:02}, but in fact, this feature
is also consistent with static magnetism that is disordered. 
At $y \! = \! 6.46$, the change in the high-field tail is accompanied by the 
appearance of a low-field tail, the origin of which is explained below. 
Similar differences between the $\mu$SR line shapes above and below the critical 
doping for the MIC are also observed in La$_{2-x}$Sr$_x$CuO$_4$ (Fig.~2). With
increasing magnetic field, the observed changes in the tails of the $x \! = \! 0.145$ 
line shape [see Figs.~2(b), 2(c) and 2(d)] are consistent with an increased density
of vortices with static electronic moments. With increasing temperature, the $\mu$SR
line shape of the $x \! = \! 0.145$ sample becomes more like that of the samples
above $x_c \! = \! 0.16$ [see Figs.~2(c), 2(e) and 2(f)], 
signifying thermal destruction of the static magnetism in and around 
the vortex cores.  

In the vortex state, the $\mu$SR signal is described by
\begin{equation}
P(t) = \sum_i \cos[\gamma_\mu B(r_i) t] \, ,
\label{eq:polarization}  
\end{equation}
where the sum is over all sites in the real-space unit cell of the VL and $B(r_i)$ 
is the local field at position $r_i \! = \! (x_i, y _i)$ with respect to the vortex 
center. Previous studies \cite{Sonier:00} showed that 
the $\mu$SR signal from high-$T_c$ superconductors is well described by Eq.~(1),
assuming a conventional phenomenological model for $B(r_i)$ and 
multiplying $P(t)$ by a Gaussian function 
$\exp(-\sigma^2 t^2 /2)$ to account
for pinning-induced disorder of the VL \cite{Brandt:88} and the static local-field 
inhomogeneity created by nuclear dipole moments. Here we find that this is not the 
case for samples on the low-doping side of the MIC. Furthermore, the simple model 
introduced in Ref.~\cite{Miller:02}, which assumed perfect AF order in the vortex cores 
commensurate with the crystal lattice, does not describe the low-field data presented 
here. Since static magnetic order is not detected in La$_{2-x}$Sr$_x$CuO$_4$
near $x \! = \! 0.145$ at $H \! < \! 30$~kOe by neutron scattering \cite{Khaykovich:05}, 
the static magnetism detected here must be disordered, 
such that the polarization function is given by
\begin{equation}
P(t) = \sum_i \exp(-\Lambda e^{-(r_i/\xi_{ab})^2} t) \cos[\gamma_\mu B(r_i) t].\
\label{eq:disaf}  
\end{equation} 
For simplicity the relaxation rate here is assumed to fall off as a function of 
radial distance $r$ from the vortex core center on the scale of the in-plane SC coherence 
length $\xi_{ab}$, and the distribution of fields at each site is assumed to be 
Lorentzian, as is often the case in SG systems.

\begin{figure}
\centering
\includegraphics[width=8.0cm]{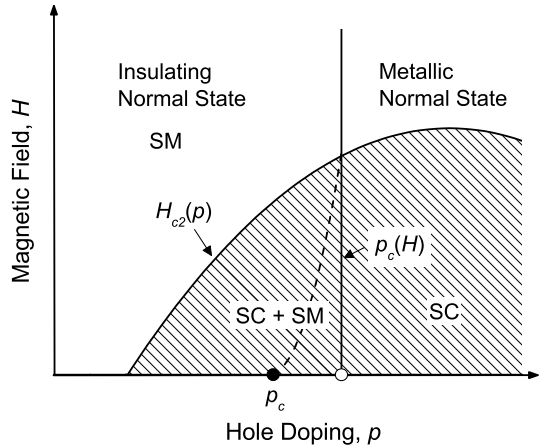}
\caption{Schematic $T \! = \! 0$~K phase diagram deduced 
from this study. The normal and SC (hatched region) phases occur above and below 
the upper critical field $H_{c2}(p)$, respectively. The solid vertical line at $p_c(H)$
represents a QPT at $H \! > \! 0$ coinciding with the low-temperature normal-state MIC. 
Below $H_{c2}(p)$, the phase boundary $p_c(H)$ separates a pure SC phase from a SC phase 
coexisting with static magnetism (SM). Immediately to the left of $p_c(H)$ the SM is 
disordered, becoming spatially uniform (and possibly ordered) above the dashed curve. 
The open circle is an `avoided' QCP, whereas the solid circle indicates 
the `true' QCP at $H \! = \! 0$.}
\label{fig3}
\end{figure} 

Figure~2(g) shows results of fits of the $\mu$SR signal for several 
samples to Eq.~(2), assuming a simple analytical solution of the Ginzburg-Landau 
equations \cite{Yaouanc:97} for $B(r_i)$. In the $x \! = \! 0.145$ and $y \! = \! 6.50$
samples at $T \! = \! 2.5$~K, the average magnetic field created by the magnetism 
at the site of a muon stopping in 
the center of a vortex core is approximately $\pm$~70~G, which is the half-width at 
half-maximum of the Lorentzian field distribution assumed in Eq.~(2). This explains 
why the appearance of magnetism in and around the vortex cores also affects the 
low-field tail of the $\mu$SR line shape. The diverging temperature dependence of 
$\Lambda$ indicates a slowing down of Cu spin fluctuations, entirely 
consistent with the approach to a second-order magnetic phase transition at 
$T \! = \! 0$~K.

We can rule out other possible origins of the observed changes in the $\mu$SR 
line shape across the MIC. First, a change in $\xi_{ab}$ would alter the location 
of the high-field `cut off', but would not change the amplitude of the high-field 
tail or introduce a low-field tail. Second, a change in symmetry of the VL affects 
the entire $\mu$SR line shape, not just the tails. In fact, if changes in 
$\xi_{ab}$ and/or VL symmetry occurred, they would be detectable in our analysis of 
the $\mu$SR line shapes \cite{Sonier:04}. Third, we consider the possibility that 
an order-to-disorder transition of the VL occurs at the critical doping for the MIC. 
A field-induced Bragg-to-vortex glass transition has been observed by $\mu$SR and 
neutron scattering in severely underdoped La$_{1.9}$Sr$_{0.1}$CuO$_4$ \cite{Divakar:04}.
The static disorder in the vortex glass phase results in a highly symmetric and 
greatly broadened $\mu$SR line shape. However, here the $\mu$SR line shapes for 
the samples on the low-doping side of the MIC are narrower than those of the 
higher-doped samples, in accordance with an increase in the magnetic penetration 
depth at lower doping. Furthermore, a transition to a vortex-glass phase is not 
supported by the increased similarity of the line shapes in Fig.~2(f). 

Our experiments establish the occurrence of a field-induced QPT below the SC dome that 
coincides with the critical doping for the MIC. At weak fields, competing 
disordered static magnetism is stabilized in and around weakly interacting vortices 
at the QPT. This confirms one of the main theoretical predictions of the modified 
phase diagram proposed by Kivelson and coworkers \cite{Kivelson:02}. The original 
theory of Ref.~\cite{Demler:01} assumed that the vortices are two-dimensional. By 
including the interlayer couplings of vortices in neighbouring CuO$_2$ layers, 
Kivelson {\it et al.} showed that a competing phase could be stabilized in nearly 
isolated vortices, thus altering the position and character of the QPT. In their 
extended theoretical model, the pure superconductor undergoes a field-induced QPT to 
a coexistence phase in which the competing order parameter (which we identify here as 
the mean squared local magnetization) is spatially inhomogeneous. With increasing field, 
stronger overlap of the magnetism around neighboring vortices may lead to a co-operative 
bulk crossover to long-range magnetic order, as is apparently the case in 
La$_{1.856}$Sr$_{0.144}$CuO$_4$ \cite{Khaykovich:05}. A key prediction of the theory of 
Ref.~\cite{Kivelson:02} is that there is an `avoided' QCP at $H \! = \! 0$ (Fig.~3). 
In other words, the QCP lies at a lower doping than one expects from the
extrapolated $H \! \rightarrow \! 0$ location of the QPT found here. 
Indeed, previous ZF-$\mu$SR measurements indicate that the onset temperature 
for static magnetism coexisting with superconductivity on a nanometer scale
in La$_{2-x}$Sr$_x$CuO$_4$ \cite{Panago:02,Niedermayer:98} and
YBa$_2$Cu$_3$O$_y$ \cite{Kiefl:89,Sanna:04}, extrapolates to zero at a hole 
concentration below the critical doping for the MIC. 
          
We thank S.A. Kivelson, R. Greene, B. Lake and E. Demler for informative discussions. 
J.E. Sonier, J.H. Brewer, R.F. Kiefl, D.A. Bonn, W.N. Hardy and R. Liang acknowledge 
support from the Canadian Institute for Advanced Research and the Natural Sciences and 
Engineering Research Council of Canada. Y. Ando acknowledges support from 
Grant-in-Aid for Science provided by the Japan Society for the Promotion of Science.


\begin{thebibliography}{xx}

\bibitem{Panago:02} C.~Panagopoulos {\it et al.}, Phys.~Rev.~B {\bf 66}, 064501 (2002).

\bibitem{Panago:05} C.~Panagopoulos, and V.~Dobrosavljevi\'{c}, Phys.~Rev.~B {\bf 72}, 014536 (2005). 

\bibitem{Khaykovich:05} B.~Khaykovich {\it et al.}, Phys.~Rev.~B {\bf 71}, 220508(R) (2005).

\bibitem{Tallon:01} J.L.~Tallon, and J.W.~Loram, Physica~C {\bf 349}, 53 (2001).

\bibitem{Ando:95}Y.~Ando {\it et al.}, Phys.~Rev.~Lett. {\bf 75}, 4662 (1995).

\bibitem{Boebinger:96} G.S.~Boebinger {\it et al.}, Phys.~Rev.~Lett. {\bf 77}, 5417 (1996).  

\bibitem{Fournier:98} P.~Fournier {\it et al.}, Phys.~Rev.~Lett. {\bf 81}, 4720 (1998).

\bibitem{Ono:00} S.~Ono {\it et al.}, Phys.~Rev.~Lett. {\bf 85}, 638 (2000).  

\bibitem{Sun:03} X.F.~Sun, S.~Komiya, J.~Takeya, and Y.~Ando, Phys.~Rev.~Lett. {\bf 90}, 117004 (2003).

\bibitem{Hawthorn:03} D.G.~Hawthorn {\it et al.}, Phys.~Rev.~Lett. {\bf 90}, 197004 (2003).

\bibitem{Sun:04} X.F.~Sun, K.~Segawa, and Y.~Ando, Phys.~Rev.~Lett. {\bf 93}, 107001 (2004).  

\bibitem{Dagan:04} Y.~Dagan {\it et al.}, Phys.~Rev.~Lett. {\bf 92}, 167001 (2004).  

\bibitem{Dagan:05} Y.~Dagan {\it et al.}, Phys.~Rev.~Lett. {\bf 94}, 057005 (2005).

\bibitem{Katano:00} S.~Katano, M.~Sato, K.~Yamada, T.~Suzuki, and T.~Fukase, 
Phys.~Rev.~B {\bf 62}, R14677 (2000).

\bibitem{Lake:02} B.~Lake {\it et al.}, Nature {\bf 415}, 299 (2002).

\bibitem{Savici:05} A.T.~Savici {\it et al.}, Phys.~Rev.~Lett. {\bf 95}, 157001 (2005).

\bibitem{Lake:01} B.~Lake {\it et al.}, Science {\bf 291}, 1759 (2001).

\bibitem{Demler:01} E.~Demler, S.~Sachdev, and Y. Zhang, Phys.~Rev.~Lett. {\bf 87}, 067202 (2001).   

\bibitem{Kivelson:02} S.A.~Kivelson, D.-H.~Lee, E.~Fradkin, and V.~Oganesyan, 
Phys.~Rev.~B {\bf 66}, 144516 (2002).

\bibitem{Sonier:00} J.E.~Sonier, J.H.~Brewer, and R.F.~Kiefl, Rev.~Mod.~Phys.
{\bf 72}, 769 (2000).

\bibitem{Miller:02} R.I.~Miller {\it et al.}, Phys.~Rev.~Lett. {\bf 88}, 137002 (2002).

\bibitem{Brandt:88} E.H.~Brandt, Phys.~Rev.~B {\bf 37}, 2349(R) (1988).

\bibitem{Yaouanc:97} A.~Yaouanc, P.~Dalmas~de~R\'{e}otier, and E.H.~Brandt, Phys.~Rev.~B 
{\bf 55}, 11107 (1997).

\bibitem{Sonier:04} J.E.~Sonier {\it et al.}, Phys.~Rev.~Lett. {\bf 93}, 017002 (2004).

\bibitem{Divakar:04} U.~Divakar {\it et al.}, Phys.~Rev.~Lett. {\bf 92}, 237004 (2004).

\bibitem{Niedermayer:98} Ch.~Niedermayer {\it et al.}, Phys.~Rev.~Lett. {\bf 80}, 3843 (1998).

\bibitem{Kiefl:89} R.F.~Kiefl {\it et al.}, Phys.~Rev.~Lett. {\bf 63}, 2136 (1989).

\bibitem{Sanna:04} S.~Sanna {\it et al.}, Phys.~Rev.~Lett. {\bf 93}, 207001 (2004).
 
\end{thebibliography}
\end{document}